\documentclass[12pt]{article}
\usepackage{amsmath}
\usepackage{graphicx,psfrag,epsf}
\usepackage{enumerate}
\usepackage{natbib}
\usepackage{url} 
\usepackage{inputenc}
\usepackage{mathtools,stmaryrd}
\usepackage{aliascnt}
\usepackage{amsthm}
\usepackage{accents}
\usepackage{natbib}
\usepackage{graphicx,psfrag,epsf}
\usepackage{amssymb}
\usepackage{mathrsfs}
\usepackage{bbm}
\usepackage{bm}
\usepackage{tikz}
\usetikzlibrary{arrows}

\usepackage{aliascnt}
\usepackage[colorlinks,citecolor=blue,linkcolor=red]{hyperref}
\usepackage[nameinlink,capitalize]{cleveref}
\usepackage{enumerate}
\usepackage{bbm}
\usepackage{bm}
\usepackage{color}
\usepackage{xargs}
\usepackage{blkarray}
\usepackage[ruled,linesnumbered]{algorithm2e}

\newcommand{\pscal}[2]{\langle #1, #2  \rangle}
\newcommandx{\norm}[2][2=]{\| #1 \|_{#2}}
\newcommandx{\nint}[2]{\lbrace #1, #2 \rbrace}
\newcommandx{\maxi}[2][2=]{\underset{#2}{\operatorname{max}}\left\{#1\right\}}
\newcommandx{\mini}[2][2=]{\underset{#2}{\operatorname{min}}\left\{#1\right\}}
\newcommandx{\argmaxi}[2][2=]{\underset{#2}{\operatorname{argmax}}\left\{#1\right\}}
\newcommandx{\argmini}[2][2=]{\underset{#2}{\operatorname{argmin}}\left\{#1\right\}}

\newcommandx{\rk}[1]{\operatorname{rk}(#1)}
\newcommandx{\prob}[1]{\mathbb{P}\left(#1 \right) }
\newcommandx{\e}{\mathrm{e}}
\newcommandx{\pe}[1]{\mathbb{E}\left[#1 \right] }
\newcommandx{\TX}{X^0}
\newcommandx{\TS}{S^0}
\newcommandx{\TL}{L^0}
\newcommandx{\Talpha}{\alpha^0}
\newcommandx{\smax}{\sigma_{\textsc{max}}}
\newcommandx{\smin}{\sigma_{\textsc{min}}}
\newcommandx{\R}{\mathbb{R}}

\usepackage[textsize=footnotesize]{todonotes} 

\makeatletter
\def\algbackskip{\hskip-\ALG@thistlm}
\def\big#1{{\hbox{$\left#1\vbox to35\p@{}\right.\n@space$}}}
\makeatother
\newcommand{\Norm}[1]{\left\lVert#1\right\rVert}

\SetKwInOut{Initialization}{Initialization}
\newcommand{\blind}{0}

\begin{document}

\def\spacingset#1{\renewcommand{\baselinestretch}%
{#1}\small\normalsize} \spacingset{1}


\if0\blind
{
  \title{\bf Imputation of mixed data with multilevel singular value decomposition}
  \author{Fran\c{c}ois Husson\hspace{.2cm}\\
    Department of statistics, Agrocampus Ouest Rennes,\\
     IRMAR, UMR CNRS 6625, France\\
    Julie Josse\hspace{.2cm}\\
    Center of Applied Mathematics, \'Ecole Polytechnique, Paris, France\\
     INRIA XPOP\\
    Balasubramanian Narasimhan\hspace{.2cm}\\
    Department of Statistics and Biomedical Data Science, Stanford University,\\
     Palo Alto, US\\
    Genevi\`{e}ve Robin\hspace{.2cm}\\
    Center of Applied Mathematics, \'Ecole Polytechnique, Paris, France \\
     INRIA XPOP\\
     and\\
     Traumabase$^{\textregistered}$ group\hspace{.2cm}\\
     H\^{o}pital Beaujon, Clichy, France}
  \maketitle
} \fi

\if1\blind
{
  \bigskip
  \bigskip
  \bigskip
  \begin{center}
    {\LARGE\bf Title}
\end{center}
  \medskip
} \fi
\bigskip
\begin{abstract}
Statistical analysis of large data sets
offers new opportunities to better understand many processes.
Yet, data accumulation often implies relaxing acquisition procedures
or compounding diverse sources. As a consequence, such data sets often contain mixed data, \textit{i.e.} both quantitative and qualitative and many missing values. 
Furthermore, aggregated data present a natural \textit{multilevel} structure, where individuals or samples are nested within different sites, such as countries or hospitals.
Imputation of multilevel data has therefore drawn some attention recently, but current solutions are not designed to handle 
mixed data, and suffer from important drawbacks such as their computational cost. In this article, we propose a single imputation method for multilevel data, which can be used to complete either quantitative, categorical or mixed data. The method is based on multilevel singular value decomposition (SVD), which consists in decomposing the variability of the data into two components, the between and within groups variability, and performing SVD on both parts. We show on a simulation study that in comparison to competitors, the method has the great advantages of handling data sets of various size, and being computationally faster. Furthermore, it is the first so far to handle mixed data. We apply the method to impute a medical data set resulting from the aggregation of several data sets coming from different hospitals. This application falls in the framework of a larger project on Trauma patients. 
To overcome obstacles associated to the aggregation of medical data, we turn to distributed computation. The method is implemented in an R package.\\

\end{abstract}

\noindent%
\vfill

\newpage
\spacingset{1.45} 
\section{Introduction}
\label{introduction}

Consider a dataset $Y\in \mathbb{R}^{n\times p}$ which is naturally the row concatenation of $K$ smaller datasets $Y_k\in \mathbb{R}^{n_k\times p}$, $k\in \{1,...,K\}$. $Y$ collects the measurements of $p$ variables across a population of $n$ individuals categorized in $K$ groups, such that the $k$-th group contains $n_k$ individuals and $\sum_{k=1}^K n_k = n$:
$$Y =\left( \begin{array}{c}
  Y_{1}\\
  \hline
  Y_{2}  \\
   \hline
  \vdots \\
  \hline
  Y_{K}\\ 
\end{array}\right)  \begin{array}{c}
  \updownarrow n_1\\
  \updownarrow n_2 \\
  \vdots \\
  \updownarrow n_K\\ 
\end{array}.$$
For a group $k\in \{1,...,K\}$, an individual of the $k$-th group $i_k\in \{1,...,n_k\}$ and a variable $j\in \{1,...,p\}$, we denote by $y_{k,i_k, j}$ the value of variable $j$ taken by individual $i_k$ in group $k$. Such structure is often called \textit{multilevel structure}, and occurs in many fields of applications. Famous examples include pupils nested within schools or patients within hospitals. Throughout this article, we focus on this latter example with a running application in public health. If some entries of $Y$ are missing, we denote by $M$ the indicator matrix of observations, with $M_{k,i_k,j} = 1$ if $y_{k,i_k,j}$ is observed and $M_{k,i_k,j} = 0$ otherwise. To handle missing values, corresponding to $M_{k,i_k,j} = 0$, a popular approach \citep{Little02} consists in imputing them, \textit{i.e.} replacing the missing entries with plausible values to get a completed data set.

To do so, several approaches have been developed, and a complete overview of state of the art multilevel imputation methods is available in \cite{Audi17}. Latest proposals have focused on handling both sporadically missing values, which correspond to some entries missing for some variables, and systematically missing values where some variables are completely unobserved in one or more groups. To take into account the hierarchical structure of the data,  most imputation methods are based on random effects regression models, such as suggested by \cite{Resche2016} and  \cite{Quar2016}. However, current solutions suffer from important gaps that deserve further development. In particular, they are not designed to handle mixed data (quantitative and categorical), struggle with large dimensions and are extremely costly in terms of computations. 

In the same time, imputation by iterative singular value decomposition (SVD) algorithms have proven excellent imputation capacities for quantitative \citep{softImpute}, qualitative \citep{Audigier2017} and mixed data \citep{Audigier2016}. This can be explained in part because they assume an underlying low-rank structure for the data which is plausible for many large data sets, as discussed in \cite{Udell17}. These methods behave particularly well compared to competitors in terms of prediction of the missing values, in particular when the number of observations is small with respect to the number of variables, and when the qualitative variables have many categories and some of them are rare. In addition, they are often competitive in terms of execution time. However, these methods are not dedicated to the multilevel data we address in this paper. The work we present here can be casted as an extension of single imputation methods based on SVD to the multilevel framework. 

The paper is organized as follows. In \Cref{model}, we start by presenting multilevel component methods to analyze quantitative, categorical and mixed data in the complete case where all entries are observed. We begin in \Cref{mult-pca} by reviewing
the multilevel simultaneous component analysis (MLSCA) of \cite{Timmerman2006MLCA}, dedicated to quantitative data, which operates by estimating principal directions of variability for both levels of variability, \textit{i.e.} for the between groups variability and for the within groups variability. Then, our first main contribution is to derive in \Cref{mult-mca} and \Cref{mult-famd} two multilevel component methods to analyze qualitative and mixed data respectively. To the best of our knowledge, we are the first to propose such methods. Our second main contribution is to propose in \Cref{missing} multilevel single imputation methods to impute categorical and mixed variables with a multilevel structure. In \Cref{simu} we show on synthetic data that our methods have smaller prediction errors than competitors when the data are generated with a multilevel model. Finally, in \Cref{trauma}, we illustrate the methods with the imputation of a large register from Paris hospitals and discuss how to distribute the computation. 
The methods  are implemented in the R \citep{R} package missMDA \citep{missMDA}.

\section{Multilevel component methods}
\label{model}

\subsection{Multilevel Principal Component Analysis (MLPCA)}
\label{mult-pca}
For sake of clarity, we start by recalling the multilevel extension of principal component analysis (PCA, \cite{Pearson:1901:PCA}) described in \cite{Timmerman2006MLCA}. Assume the data set $Y$ contains quantitative variables only. The measured values can be decomposed, for a group $k\in \{1,...,K\}$, an individual $i_k\in \{1,...,n_k\}$ in the $k$-th group and a variable $j\in \{1,...,p\}$, as
\begin{equation*}
\begin{aligned}
& y_{k,i_k,j} = \underbrace{y_{.,.,j}}_{\text{offset}} + \underbrace{y_{k,.,j}-y_{.,.,j}}_{\text{between}} + \underbrace{y_{k,i_k,j}-y_{.,k,j}}_{\text{within}}.
\end{aligned}
\end{equation*}
Here, $$y_{.,.,j} = \frac{1}{n}\sum_{k=1}^K\sum_{i_k=1}^{n_k}y_{k,i_k,j}$$ is the overall mean of variable $j$ and $$y_{k,.,j} = \frac{1}{n_k}\sum_{i_k=1}^{n_k}y_{k,i_k,j}$$ is the mean of variable $j$ among individuals of group $k$. Then, $(y_{k,.,j}-y_{.,.,j})$ is the deviation of group $k$ to the overall mean of variable $j$, and $(y_{k,i_k,j}-y_{k,.,j})$ is the deviation of individual $i_k$ to the mean of variable $j$ in group $k$. Written in matrix form, this gives
$$Y =1_{n}m^{\top} + Y_b + Y_w,$$
where $1_{n}$ is the $n\times 1$ vector of ones and $m$ is the $p \times 1$ vector containing the overall means of the $p$ variables, $Y_b$ contains the variable means per group minus the overall means, and $Y_w$ contains the residuals. Similarly to what is done in analysis of variance, we can split the sum of squares for each variable $j$ as
\begin{equation*}
\sum_{k=1}^K\sum_{i_k=1}^{n_k}y_{k,i_k,j}^2 = \sum_{k=1}^Kn_ky_{.,.,j}^2 + \sum_{k=1}^{K}n_k(y_{k,.,j}-y_{.,.,j})^2 +\sum_{k=1}^K\sum_{i_k=1}^{n_k}(y_{k,i_k,j}-y_{k,.,j})^2.
\end{equation*}
In the classical framework where there is no multilevel structure, PCA yields the best fixed rank estimator of $Y$ in terms of the least squares criterion. The multilevel extension naturally leads, for $(k,i_k,j)\in \{1,\ldots, K\}\times \{1,\ldots, n_k\} \times \{1,\ldots, p\}$, to modelling the offsets, the between and within terms separately by explaining as well as possible both the between and within sum of squares. Therefore, multilevel PCA (MLPCA) consists in assuming  two low-rank models, for the between matrix $Y_b = (y_{k,.,j}-y_{.,.,j})_{k,j}$ - that we approximate by a matrix of rank $Q_b$, and for the within matrix $Y_w = (y_{k,i_k,j}-y_{k,.,j})_{k,i_k,j}$ - that we approximate by a matrix of rank $Q_w$. This yields the following decomposition: 
\begin{equation} \label{eq:mlscamodel}
Y=1_{n}m^{\top}+F_{b}V_b^{\top}+F_{w}V_{w}^{\top} + E.
\end{equation}
 $F_{b}$ is the matrix of size $n\times Q_b$ containing the between component scores
\begin{equation}
\label{u-struct}
F_b =\left( \begin{array}{c}
  F_{b,1}\\
  \hline
  F_{b,2}  \\
   \hline
  \vdots \\
  \hline
  F_{b,K}\\ 
\end{array}\right),
\end{equation}
where for all $k\in\{1,\ldots,K\}$, $F_{b,k}$ is row-wise constant, with $f_{b,k}$ repeated on every row. Let $I_k\in\{0,1\}^n$ be the indicator vector of group $k$ such that the $i$-th entry $I_{k,i} =1$ if individual $i$ belongs to group $k$ and $0$ otherwise. Representation \eqref{u-struct} is equivalent to 
$$F_b = \sum_{k=1}^K I_kf_{b,k}^{\top}.$$
$V_b$ is the $p \times Q_b$ between loadings matrix, $F_{w}$ $(n \times Q_{w})$ denotes the within component scores, and finally $V_{w}$ $(p \times Q_{w})$ denotes the within loadings matrix, and $E$ $(n \times p)$ denotes the matrix of residuals. Note that in this model,  the within loadings matrix $V_{w}$ is constrained to be constant across groups.
Model \eqref{eq:mlscamodel} is called multilevel simultaneous component analysis (MLSCA) in \cite{Timmerman2006MLCA}. We keep the name MLPCA for simplicity. 

In terms of interpretation, the low rank structure on the between part implies that there are dimensions of variability to describe the hospitals: for instance the first dimension could oppose hospitals  that resort to a large extent to pelvic and chest X-ray to hospitals where those examinations are not usually performed. The low rank structure on the within part  implies that there are dimensions of variability to describe the patients: for instance the first dimension opposes patients with a head trauma (taking specific values for variables related to head trauma) to other patients.
The constraint that the within loading matrix is the same across hospitals means that this dimension is the same from one hospital to the other but the strength of the dimension, \textit{i.e.} the variability of patients on the dimension, can differ from one group to the other.
This constraint also leads to less parameters to estimate.

The model is fitted by solving the least squares problem with respect to the parameter $\Theta = (m,F_{b},V_b,F_{w},V_{w})$:

\begin{equation} 
\label{eq:lsmlsca}
\begin{aligned}
\hat \Theta \in &\operatorname{argmin}_{\Theta}\quad\left\|Y-\left(1_{n}m^{\top}+F_{b}V_b^{\top}+F_{w}V_{w}^{\top}\right) \right\|^2,\\
& \text{such that}\quad F_b = \sum_{k=1}^K I_kf_{b,k}^{\top}\text{, }\sum_{k=1}^K n_kf_{b,k} = 0_{Q_b}\text{, }1_{n}^{\top}F_{w}=0_{Q_w},
\end{aligned} 
\end{equation}
where the last two constraints serve for identifiability. 
The problem is separable, and the solution is obtained in \cite{Timmerman2006MLCA} by computing the variables means to estimate $m$, the matrix of means per group centered by the overall mean $Y_b$ and the within matrix $Y_w$ of the data centered per group. Then, truncated SVD of $Y_b=U_b\Lambda_b^{1/2}V_b^{\top}$ at rank $Q_b$ and of $Y_w=U_w\Lambda_w^{1/2}V_w^{\top}$ at rank $Q_w$ are performed to estimate the parameters $(F_b=U_b\Lambda_b^{1/2},V_b, F_w=U_w\Lambda_w^{1/2}, V_w)$.
Such a solution is in agreement with the rationale of performing an SVD on the matrix of means per group to study the differences between groups and a SVD of the matrix centered by groups to study the differences between patients after discarding the hospital effects. 

\subsection{Multilevel Multiple Correspondence Analysis (MLMCA)}
\label{mult-mca}

We now propose a new counterpart of MLPCA to analyse categorical variables. Our method is based on multiple correspondence analysis (MCA, \cite{Green06,paghuss2010da}), that we extend to handle multilevel structures.
MCA is considered to be the counterpart of PCA for categorical data analysis, and has been successfully applied in many fields of applications, such as survey data analysis, to visualize associations between categories. More precisely, categorical data are coded as a complete disjunctive table $Z$ where all categories of all variables are represented as indicator vectors. In other words $z_{ic}=1$ if individual $i$ takes the category $c$ and $0$ otherwise.
 For example, if there are $p=2$ variables with 2 and 3 levels respectively, we have the following equivalent codings:
\begin{equation*}\label{eq:catTables}
  Y = \begin{pmatrix}
    1 & 1\\
    2 & 3\\
    1 & 2\\
    2 & 3\\
    2 & 2\\
    2 & 2
  \end{pmatrix}
  \;\;\; \Longleftrightarrow \;\;\;
  Z= \begin{pmatrix}
    1 & 0 & & 1 & 0 & 0\\
    0 & 1 & & 0 & 0 & 1\\
    1 & 0 & & 0 & 1 & 0\\
    0 & 1 & & 0 & 0 & 1\\
    0 & 1 & & 0 & 1 & 0\\
    0 & 1 & & 0 & 1 & 0
  \end{pmatrix}.
\end{equation*}
For $1\leq j\leq p$ we denote by $C_j$ the number of categories of variable $j$, and $C = \sum_{j=1}^p C_j$ the total number of categories. For $1\leq c\leq C$, $Z_{,c}$ is the $c$-th column of $Z$ corresponding to the indicator of category $c$. We define $\pi_c=n^{-1}1_n^{\top}Z_{,c}$ the proportion of observations in category $c$, $\pi=(\pi_1,\ldots, \pi_C)^{\top}$ and $D_{\pi}$ the $C\times C$ diagonal matrix with $\pi$ on its diagonal. Multiple correspondence analysis (MCA) is defined as the SVD of the matrix
\begin{eqnarray} \label{eq:mca}
 A = \frac{1}{np}\left(Z-1_n \pi^{\top}\right) {D_{\pi}}^ {-1/2}.
\end{eqnarray}
This specific transformation endows MCA with many properties: the distances between the rows and columns in the transformed matrix $A$ coincide with the chi-squared distances, the first principal component (the scores) is the quantitative variable most related to the categorical variables in the sense of the $\eta^2$ coefficient of analysis of variance \cite[Section 3]{paghuss2010da}. This latter property justifies why  MCA is considered as the equivalent of PCA for categorical data.

We introduce the following strategy for multilevel MCA (MLMCA). From the indicator matrix of dummy variables $Z$, we start by defining a between part and a within part. MCA, in the sense of the SVD of a transformed matrix \eqref{eq:mca},  will then be applied on each part. For $k\in\{1,...,K\}$, define $Z_k$ the sub-matrix of $Z$ containing all categories and the rows corresponding to individuals of group $k$. The between part is defined block-wise as the mean of the indicator matrix per group $k$ with the following $n_k\times p$ matrices,  stacked below one another:
\begin{equation*}
Z_{b,k} =n_k^{-1}1_{n_k}1_{n_k}^{\top}Z_k.
\end{equation*}
The entries of $Z_{b,k}$ contain the proportion of observations taking each category in group $k$ $(n_{c_{k}}/ n_k)$ (for instance the proportion of individuals carrying some disease in a particular hospital). Finally
\begin{equation*}
Z_b =\left( \begin{array}{c}
  Z_{b,1}\\
  \hline
  Z_{b,2}  \\
   \hline
  \vdots \\
  \hline
  Z_{b,K}\\ 
\end{array}\right).
\end{equation*}
MCA  \eqref{eq:mca} is afterwards applied to the fuzzy indicator matrix $Z_b$, \textit{i.e.} SVD is applied to
\begin{equation*}
(Z_b-1_n \pi^{\top})D_{\pi}^{-1/2}.
\end{equation*}
This results in obtaining between component scores $F_b\in \mathbb{R}^{n\times Q_b}$ and between loadings $V_b\in \mathbb{R}^{n\times Q_b}$. The estimated between matrix is then $\hat Z_b=F_b V_{b}^{\top} D_{\pi}^{1/2} + 1_n \pi^{\top}$.
As for the within part, MCA is applied to the data where the between part has been swept out, \textit{i.e.} SVD is applied to the following matrix: 
\begin{eqnarray} \label{eq:withinmca}
\left(Z- Z_b\right)D_{\pi}^{-1/2}.
\end{eqnarray}
Weighting by the inverse square root of the margins of the categories implies that more weight is given to categories which are rare over all groups (for instance a rare disease). 
We obtain within component scores $F_w\in \mathbb{R}^{n\times Q_w}$, within loadings $V_w\in \mathbb{R}^{n\times Q_w}$, and the estimated within matrix $ \hat Z_w=F_w V_{w}^{\top}D_{\pi}^ {1/2}$. \\
Finally, we estimate $Z$ by $\hat{Z}= \hat Z_b+\hat Z_w$.
As with MCA \citep{Josse2012}, the reconstructed fuzzy indicator matrix $\hat Z = \hat Z_b + \hat Z_w$ has the property that the sum of values for one individual and one variable is equal to one. Consequently, the estimated values can be considered as degrees of membership to the categories.  This property will prove useful for the imputation. 

\paragraph{Remark} Another approach to define MLMCA would have been to directly apply MLPCA on the matrix $A$ \eqref{eq:mca}. It turns out that the two strategies are equivalent  which strengthens this definition of Multilevel MCA.

\subsection{Multilevel Factorial Analysis of Mixed Data (MLFAMD)}
\label{mult-famd}
Consider now a mixed data set $Y = (Y_q, Y_c)$, where $Y_q$ is a submatrix containing $p_q$ quantitative variables, and $Y_c$ a submatrix containing $p_c$ categories:
\begin{equation*}\label{eq:mixedTables}
  Y =  \big(\underbrace{\begin{matrix}
     0.3 & -3.4 & 0.1\\
     1.4 & 0.4 & -2.8\\
     9.2 & 1.8 & 7.1
  \end{matrix}}_{Y_q}
  \;\;\; \;\;\;
  \underbrace{\begin{matrix}
    0 & 1&  & 0 & 1 & 0 \\
    1 & 0&  & 0 & 0 & 1\\
    0 &1 &  & 1 & 0 & 1
  \end{matrix}}_{Y_c} \big).
\end{equation*}
In the same flavour, we define a multilevel method for mixed data by extending a counterpart of PCA for mixed data, namely factorial analysis for mixed data (FAMD), presented in \cite{Page14}. FAMD consists in transforming the categorical variables as in MCA \eqref{eq:mca} and concatenating them with the quantitative variables. Then, each quantitative variable is standardized (centered and divided by its standard deviation).
Finally, SVD is applied to this weighted matrix. 
This specific weighting ensures that all quantitative and categorical variables play the same role in the analysis. More precisely, the principal components, denoted $F_q$ for $q=1, ..., Q$ maximize the link between the quantitative  and categorical variables in the following sense:
$$ {F}_q =\operatorname*{arg\,max}_{{F}_q \in {\mathbb R}^{n}} \sum_{j=1}^{p_{q}} r^2(F_q,Y_j)+\sum_{jc=1}^{p_{c}}\eta^2(F_q,Y_{jc}) ,$$
with the constraint that ${F}_q$  is orthogonal to ${F}_{q^{\prime}}$ for all $q'<q$ and 
with $Y_j$ being the variable $j$, $r^2$ the square of the correlation coefficient and $\eta^2$ the square of the correlation ratio.  This formulation highlights that FAMD can be seen as the counterpart of PCA for mixed data. More details about the method are given in \cite{Page14}. 

The extension to a multilevel structure, named MLFAMD, is now straightforward following what is done for MCA and categorical data in the previous section. Denote $C$ the number of categories, $\pi\in(0,1)^C$ the vector of categories proportions and $D_{\pi}$ the $C\times C$ diagonal matrix containing $\pi$ on its diagonal. Denote $m\in\mathbb{R}^{p_q}$ the vector of means of the quantitative variables, and $\Sigma\in\mathbb{R}^{p_q\times p_q}$ the diagonal matrix containing the standard deviations of $Y_q$. MLFAMD consists in doing the following transformations.
\begin{equation}
\label{eq:famd-transfo}
\begin{aligned}
W \in \mathbb{R}^{n\times (p_q+p_c)}\leftarrow \Bigg((Y_q-1_nm^{\top})\Sigma^{-1}, \frac{1}{np}\left(Y_c-1_n \pi^{\top}\right) {D_{\pi}}^ {-1/2}\Bigg).
\end{aligned}
\end{equation}
Then, multilevel SVD is performed on the matrix $W$. This boils down to computing the between and within part, and performing SVD on both separately:
\begin{equation}
\label{eq:W}
W_b = \sum_{k=1}^K1_{n_k}1_{n_k}^{\top}W,\quad W_w = W - W_b.
\end{equation}

\section{Multilevel imputation}
\label{missing}

\subsection{Imputation with MLPCA}
\label{mult-pca-mis}

We now focus on the case where some values in $Y$ are missing. Recall that $M$ is the $n\times p$ indicator matrix of observations with $M_{k,i_k,j} = 1$ if $y_{k,i_k,j}$ is observed and $M_{k,i_k,j} = 0$ otherwise. 
 We denote by $M_k$ the restriction of matrix $M$ to the rows belonging to group $k\in\{1,...,K\}$. Consider a Missing (Completely) At Random (M(C)AR) setting \citep{Little02} where the process that generated the missing values can be ignored. To impute the missing values using the multilevel model \eqref{eq:mlscamodel}, we need to estimate its parameters from the incomplete data. This can be done through low rank matrix estimation for incomplete data sets \citep{softImpute} by weighting the least
squares criterion \eqref{eq:lsmlsca} with $\{0,1\}$ weights indicating the observed entries.
Let $\Theta = (m,F_{b},V_b,F_{w},V_{w})$, the optimization problem is the following with $\odot$ denoting the Hadamard product:
\begin{equation} \label{eq:wlsmlsca}
\begin{aligned}
\Theta \in &\operatorname{argmin}_{\Theta}\quad\left\|M\odot\left(Y-(1_{n}m^{\top}+F_{b}V_b^{\top}+F_{w}V_{w}^{\top})\right) \right\|_2^2\\
& \text{such that}\quad F_b = \sum_{k=1}^K I_kf_{b,k}^{\top}\text{, }\sum_{k=1}^K n_kf_{b,k} = 0_{Q_b}\text{, }1_{n}^{\top}F_{w}=0_{Q_w}.
\end{aligned}
\end{equation}
In \cite{Josse2013MLSCA}, the authors solved such a program  using an iterative imputation algorithm. Note that the aim in \cite{Josse2013MLSCA} was to perform MLPCA with missing values, \textit{i.e.} to estimate the parameters despite the missing values, and not to impute multilevel data. The distinction may appear tenuous as the algorithm involves an underlying imputation of the missing entries, but the quality of this imputation was never evaluated in itself. Let $\hat m^0$ be the mean vector of the non-missing entries. The algorithm works iteratively as described in Algorithm \ref{algo:it-PCA}.

\begin{figure}[hbtp]
\centering
\begin{minipage}{.7\linewidth}
\begin{algorithm}[H]
\begin{enumerate}
\item[0.] \textbf{Initialize} missing values: $\hat{Y} = Y\odot M + 1_n \hat{m}^{0\top}\odot(1_n1_p^{\top}-M)$.
\item Estimate $F_{b},V_b,F_{w},V_{w}$ with multilevel PCA \eqref{eq:lsmlsca};
\item Impute  $Y = Y\odot M + (1_nm^{\top}+F_bV_b^{\top}+F_wV_w^{\top})\odot(1_n1_p^{\top}-M)$;
\item Update means $m = n^{-1}1_n^{\top} Y$.
\end{enumerate}
\textbf{Repeat} steps 1, 2, 3 until empirical stabilization of the prediction.
\caption{Iterative MLPCA}
\label{algo:it-PCA}
 \end{algorithm}
\end{minipage}
\end{figure}
 Such an algorithm starts by
replacing the missing values by initial values (for example the mean of the non-missing entries),
then the estimator (here MLPCA) is computed on the completed matrix and the predicted values of the
missing entries are updated using the values given by the new estimation. The two steps of
imputation an estimation are repeated until empirical stabilization of the prediction.

\begin{figure}[hbtp]
  \centering
  \begin{minipage}{.7\linewidth}
    \begin{algorithm}[H]
      \SetAlgoLined
      \KwData{$Y=(Y_{\text{obs}},Y_{\text{mis}})\in \mathbb{R}^{n\times p}$, $Q_b$, $Q_w$}
            \Initialization{
      $\hat{m}^0$  be the mean vector of the non-missing entries\\
      \For{$(i,j)\in\nint{1}{n}\times\nint{1}{p}$}{
      \If{$M_{ij} = 0$}{
      	$Y_{ij}\leftarrow \hat m^0_j$
      }
      }
      }
      \Repeat{convergence}{
      	\textbf{Estimation of the between structure}\\
      	$Y_b=\sum_{k=1}^Kn_k^{-1}I_k\left(1_{n_k}^{\top}Y_k - \hat{m}^{0\top}\right) $\\
		$Y_b=FV^{\top}$\text{ (SVD)}\\
		$F_b\leftarrow F[,1:Q_b]$; $V_b\leftarrow V[,1:Q_b]$\\
      	$\hat{Y}_b = F_bV_b'$\\
      	
      	\textbf{Estimation of the within structure}\\
      	$Y_w=Y-1_n\hat{m}^{\top}-Y_b$\\
		$Y_w=FV^{\top}$ \text{ (SVD)}\\
		$F_w\leftarrow F[,1:Q_w]$; $V_w\leftarrow V[,1:Q_w]$\\
      	$\hat{Y}_w = F_wV_w'$\\
      	
      	\textbf{Imputation of the missing values}\\
      	$\hat{Y} = 1_n\hat{m}^{\top}\hat{Y}_b+\hat{Y}_w$\\
      	$Y \leftarrow M\odot Y + (1_n1_p^{\top}-M)\odot\hat Y$\\
      	$\hat{m} = n^{-1}1_n^{\top}Y$
      	
      }
      \caption{Iterative MLPCA (detailed)}
      \label{dist-pca-mult}
    \end{algorithm}
  \end{minipage}
\end{figure}
The detailed algorithm for iterative MLPCA with missing values is given in Algorithm \ref{dist-pca-mult}. 
In the end, it outputs both the between and within scores and loadings obtained from the incomplete dataset, and a dataset imputed using the MLPCA model \eqref{eq:mlscamodel}. 
Thus, it is a single imputation method \citep{Schafer97, Little02} which takes into account the multilevel structure of the data.
Note also that the algorithm corresponds to an expectation-maximization (EM) algorithm  of the multilevel model \eqref{eq:mlscamodel} assuming gaussian noise.
To prevent overfitting, the SVD step is replaced by regularized SVD, \textit{i.e.} where the singular values are shrunk, as described in \Cref{sec:imp-mca-famd}. This type of regularization is classical in SVD based methods \citep{verbanck2013regularised, denoiser}

\subsection{Imputation with MLMCA and MLFAMD}
\label{sec:imp-mca-famd}
Based on Algorithm \ref{dist-pca-mult} for imputation of multilevel quantitative data, we define two iterative imputation algorithms for multilevel MCA and multilevel FAMD. They are sketched together in Algorithm \ref{algo:imp-mca-famd}.
\begin{figure}[hbtp]
\centering
\begin{minipage}{.7\linewidth}
\begin{algorithm}[H]
\begin{enumerate}
\item[(0)]\textbf{Initialization} 
\begin{enumerate}
\item[(a)] Initialize missing values: mean imputation for quantitative data, proportion imputation for dummy variables.
\item[(b)] Compute weights, standard deviations and column margins.\end{enumerate}
\item[(1)]\textbf{Repeat until convergence:}
\begin{enumerate}
\item[(a)] Estimate parameters (with MLFAMD or MLMCA)
\item[(b)] Impute the missing entries with fitted values 
\item[(c)] Update means, standard deviations, column margins.\footnote{After each imputations, the means and standard deviations are modified. Hence we need to recenter and rescale the data.
} \end{enumerate}
\end{enumerate}
\caption{Iterative MLMCA and iterative MLFAMD}
\label{algo:imp-mca-famd}
 \end{algorithm}
\end{minipage}
\end{figure}
Note that we implemented an accelerated version of the algorithm where the between and the within parts are not updated simultaneously but one at a time. This corresponds to a generalized EM step, where the least-squares criterion is decreased at every iteration of the algorithm, but not entirely minimized.\\
Note also that these methods require to select two parameters: the number of between and within components $Q_b$ and $Q_w$. Furthermore, they must be selected from an incomplete data set. This is far from trivial, especially in the case of categorical variables. In fact, even in the complete case and without multilevel structure, not many options are available. Consequently, we advocate the use of cross-validation to select these components.\\
Furthermore, to prevent overfitting we actually perform a regularized SVD where singular values are shrunk. Let $\lambda_l$, $1\leq l\leq Q_b$, and $\nu_q$, $1\leq q\leq Q_w$, be the ordered singular values of $W_b$ and $W_w$, defined in \eqref{eq:W}, respectively. Let $\hat{\sigma}_b^2=1/(K-Q_b)\sum_{s=Q_b+1}^K \lambda_s$ and $\hat{\sigma}_w^2=1/(p-Q_w)\sum_{s=Q_w+1}^p \nu_s$. We shrink the singular values as follows: 
$$\left(\lambda_1,\ldots, \lambda_{Q_b}\right) \leftarrow \left(\frac{\lambda_1-\hat{\sigma}_b^2}{\sqrt{\lambda_1}},\ldots, \frac{\lambda_{Q_b}-\hat{\sigma}_b^2}{\sqrt{\lambda_{Q_b}}}\right),$$
 $$\left(\nu_1,\ldots, \nu_{Q_w}\right)\leftarrow\left(\frac{\nu_1-\hat{\sigma}_w^2}{\sqrt{\nu_1}},\ldots, \frac{\nu_{Q_w}-\hat{\sigma}_w^2}{\sqrt{\nu_{Q_w}}}\right).$$
Finally, the algorithms we present in this paper can be implemented in parallel across groups, providing that groups share their mean values, standard deviations, sample sizes, and right singular vectors. The procedure to distribute the computation is described in \Cref{sec:appDistSVD}.  Such a procedure is interesting in the framework of the medical application described in \Cref{trauma} as it allows each hospital to keep their data on site while benefiting from other hospitals data for the imputation. 

\section{Simulation study}
\label{simu}

\subsection{Imputation of multilevel quantitative data}

We conducted a comparative simulation study to contrast the performances of the multilevel imputation with PCA (MLPCA) to other single imputation methods, namely
\begin{enumerate}
\item mean imputation which consists in imputing by the mean of each variable, used as a benchmark method;
\item a separate PCA imputation where each group is imputed independently, using the R package missMDA  \citep{missMDA};
\item a global imputation by PCA (which ignores the multilevel structure and the group variable) using the R package missMDA \citep{missMDA};
\item imputation with iterative conditional random effects regression models as implemented in the R package mice \citep{Buur12}; 
\item imputation by a joint model based on random effects models as implemented in the R package jomo \citep{jomo};
\item imputation with iterative random forest (RF) as implemented in the R package missForest,  \citep{Buhlmann12}. The group variable is included for the imputation. 
\end{enumerate}
Note that methods 4 and 5 are considered as the references to impute multilevel quantitative data \citep{Audi17}. However, these methods are defined as multiple imputation methods and used the imputed data as an intermediary to do statistical inference with missing values. Here, we compute the mean over 100 multiple imputed data to get one single imputed dataset.\\
The imputation based on random forests can handle mixed variables and is known to be a very powerful tool for imputation. 
It is not specifically designed to handle a multilevel structure, but is expected to perform well in such a hierarchical setting. Indeed, random forests can account for interactions between variables, and therefore in particular for interactions between the categorical variable indicating the group and the other variables. This is another way of handling the multilevel structure. In the same way, even though we focus here on quantitative variables, we also added imputation method for mixed data with FAMD \citep{Audigier2016}, where the group membership is used as a categorical variable. This allows to take into account the hierarchical structure of the data.\\

We first simulate data according to the multilevel model \eqref{eq:mlscamodel} with Gaussian noise and set the number of between and within components to 2. For MLPCA, global PCA and FAMD, we select the number of components resulting in the smallest errors. This corresponded to $Q_b=2$ and $Q_w=2$ for MLPCA, and to 4 dimensions for the global PCA and global FAMD. We use default parameters for the other methods.  
We start with $n_k=20$ observations per group $k$ and we vary the number of groups $K$ $(3, 5)$, the number of variables $J$ (5, 10, 30),  the intensity of the noise ($\sigma=1, 2$) and the percentage of missing values (10\%, 20\%, 30\%, 40\%), which are missing completely at random (MCAR). 
The detail is available in the associated code provided as supplementary material. 
We then compute the mean squared error (MSE) of prediction, and repeat the process 100 times. 
\begin{figure}[hbtp]  
\centering
 \includegraphics[width=10cm]{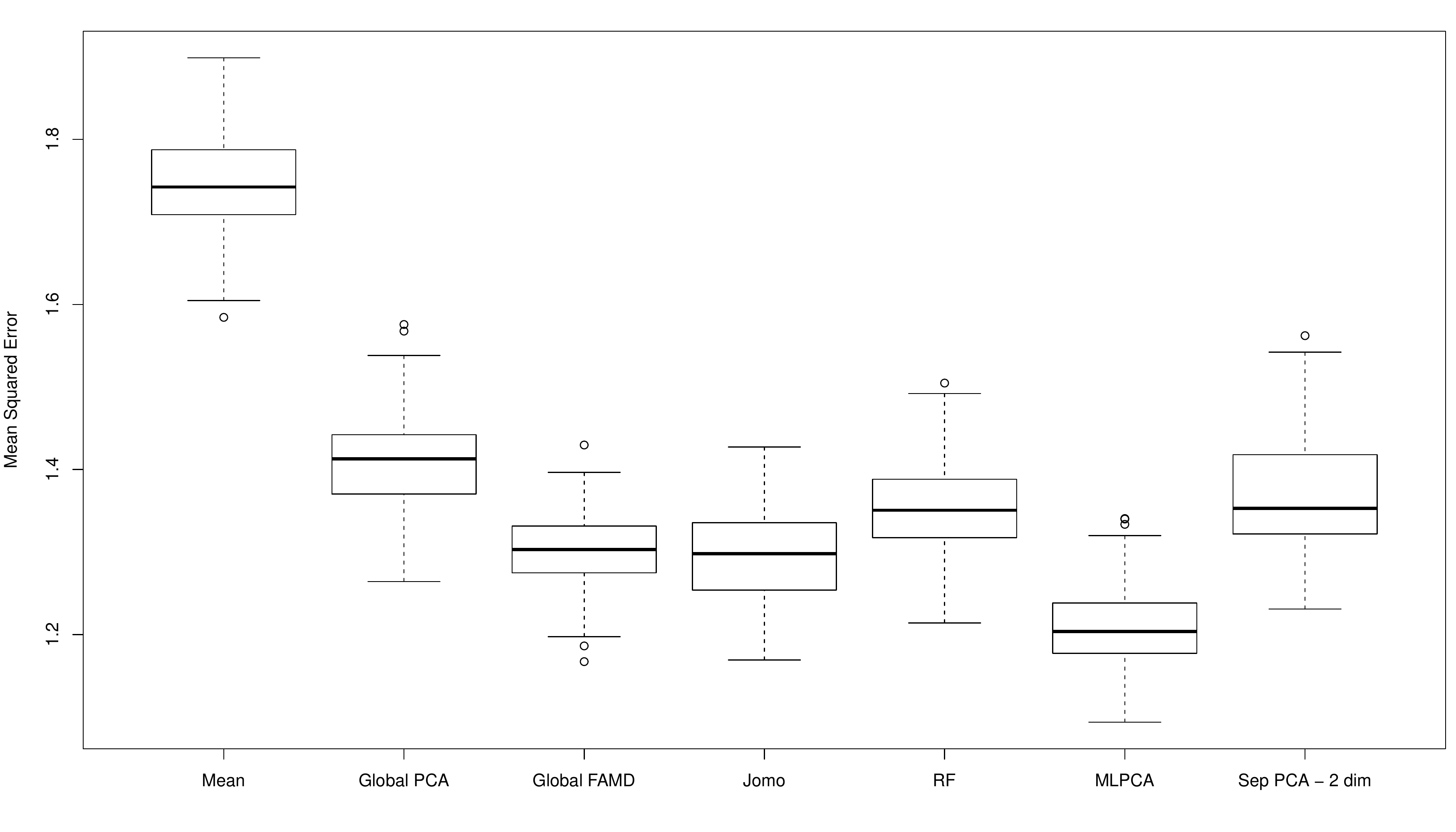} \caption{MSE of prediction for a data with $J=10$ variables,  $K=5$ groups, $n_k=20$ observations per group and  30\% of missing values completely at random. MLPCA is performed with $Q_b=2$ and $Q_w=2$.
\label{fig:summary_left}}
\end{figure}
Figure \ref{fig:summary_left} is representative of many results where multilevel imputation MLPCA improves both on global PCA imputation and separate PCA imputation but also on competitors. 
We have not included the results from the package mice as, using the default parameters, we encountered too many errors. It may be explained by the size of the data set, as the method does not behave well when there are not too many variables. More tuning is surely required to use the mice package seamlessly.\\

\begin{figure}[hbtp]  
\centering
 \includegraphics[width=6cm]{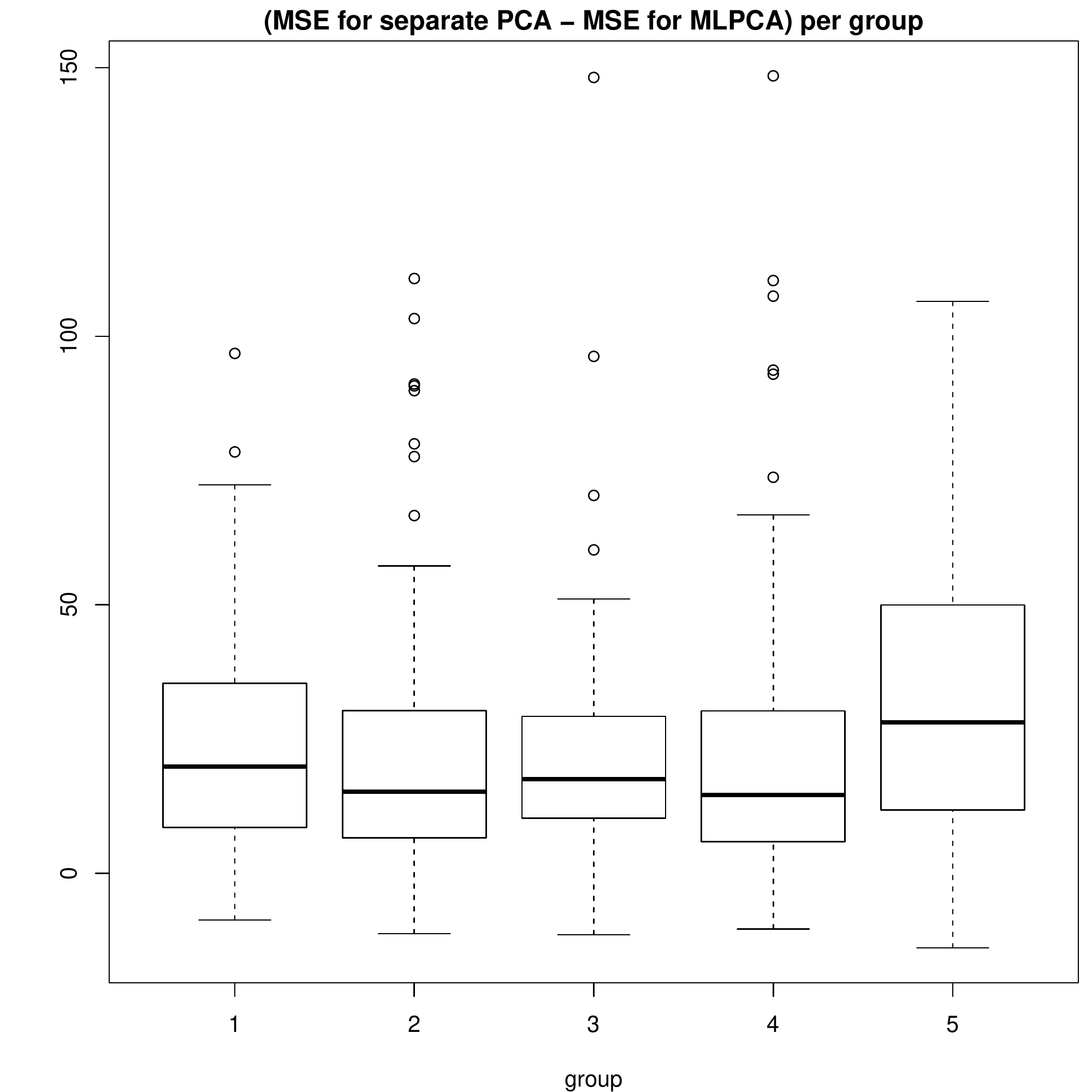} \caption{Difference between MSE obtained with separate PCA and with MLPCA for each group.
\label{fig:summary_right}}
\end{figure}

We summarize here our main findings with respect to all the simulations carried out. 
Imputations with random forests and FAMD often perform similarly with a slight advantage for FAMD especially when the percentage of missing values is large. Imputation with jomo encounters many difficulties when the number of variables increases as well as when the noise increases.  Finally imputation based on separate PCA collapses when the percentage of missing values increases and/or the number of observations per group decreases, which is not surprising as it operates on the smaller group data sets.  The multilevel imputation is always the most accurate. This is expected (but still reassuring) as the data are simulated according to a multilevel model.  
We also simulated data without a multilevel structure, \textit{i.e.} with one single group containing all individuals, and the performances of multilevel PCA are only slightly lower than those of global PCA. \\
All the methods have of course their strengths and weaknesses, and the properties of an imputation method depend on its inherent characteristics: an imputation method based on low rank assumption and linear relationships provides good prediction for data with strong linear relationships contrary to imputation using random forests which are designed for non-linear relationships.\\
 However, we observe that imputation with random forests breaks down for small sample sizes in missing at random (MAR) cases, because extrapolation and
prediction outside the range of the data seems difficult with random forests.
Since the structure of the data  is not known in advance, one could use cross-validation and select the method which best predicts the removed entries.
Figure \ref{fig:summary_right} represents the differences, for each group, between imputing with a separate PCA and with MLPCA. 
The improvement of a multilevel imputation over a separate imputation differs from one study to the other but still groups have interest in using a multilevel imputation. Indeed, the results presented in Figure \ref{fig:summary_right} reveal that in terms of predicting the missing entries, multilevel PCA yields better results that separate PCA for every group, thus showing that as far as imputation is concerned, all groups benefit from participating in the study. This justifies the use of distributed multilevel methods in contexts where there are confidentiality issues at stake, by quantifying how much the different centers gain in terms of imputation accuracy, as further discussed in \Cref{trauma}.

\subsection{Imputation of multilevel mixed data}

To simulate mixed data, we use the same design as for quantitative variables but cut some of the variables into categories. We vary the same parameters as for the quantitative variables but also the ratio of the number of quantitative over the number of categorical variables. 
Note that the methods implemented in the packages mice and jomo can handle mixed data when categorical variables are binary, but not when variables have more than two categories. This is why they are not included in the simulations.  
The global FAMD imputation is performed with 2, 4 and 6 dimensions whereas we also vary $Q_b$ and $Q_w$ for the multilevel method between 2 and 4. We display only the number of components which resulted in the lowest prediction error for each of the methods concerned.
\begin{figure}[hbtp]  
\centering
 \includegraphics[width=8cm]{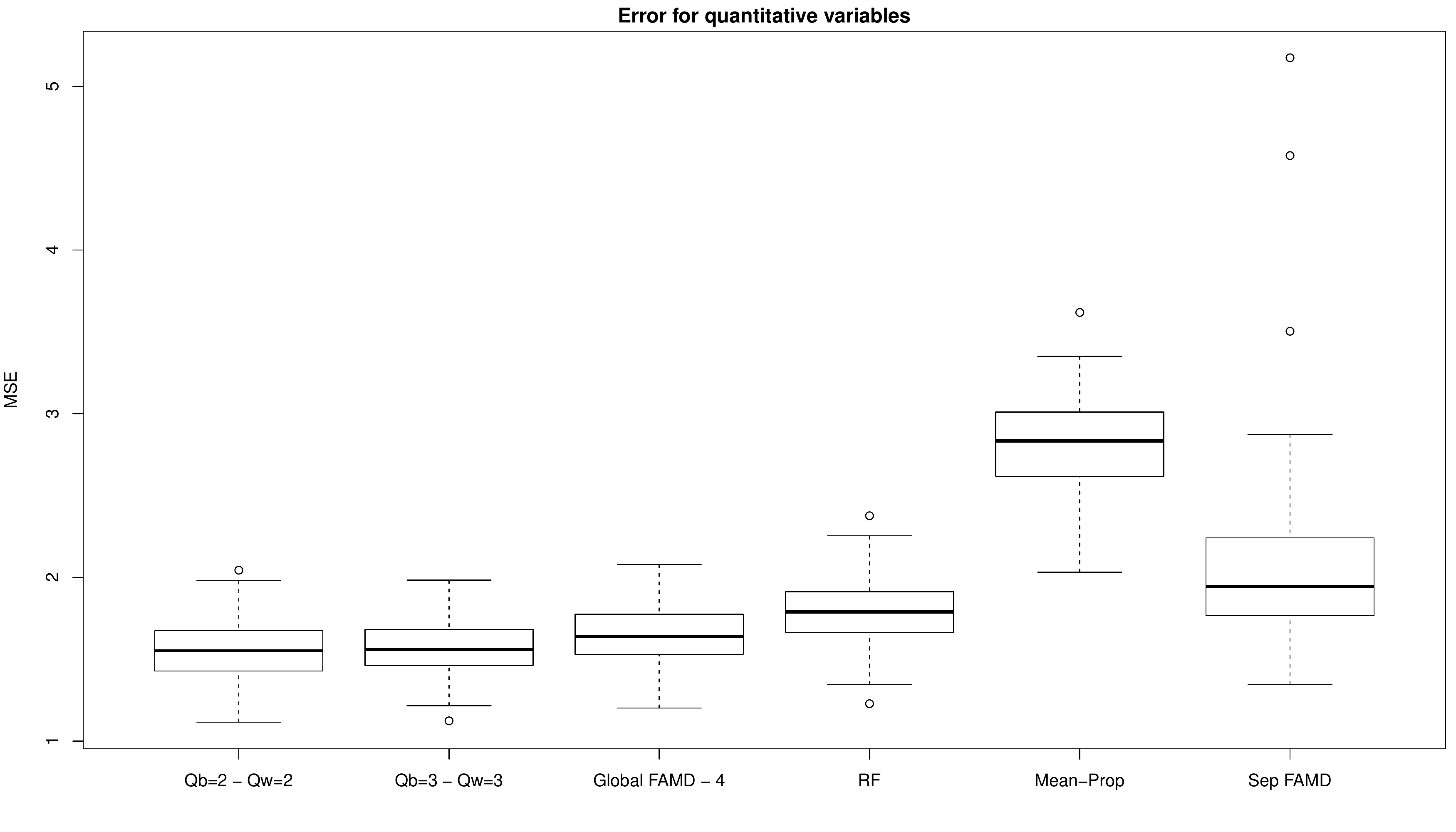} \includegraphics[width=8cm]{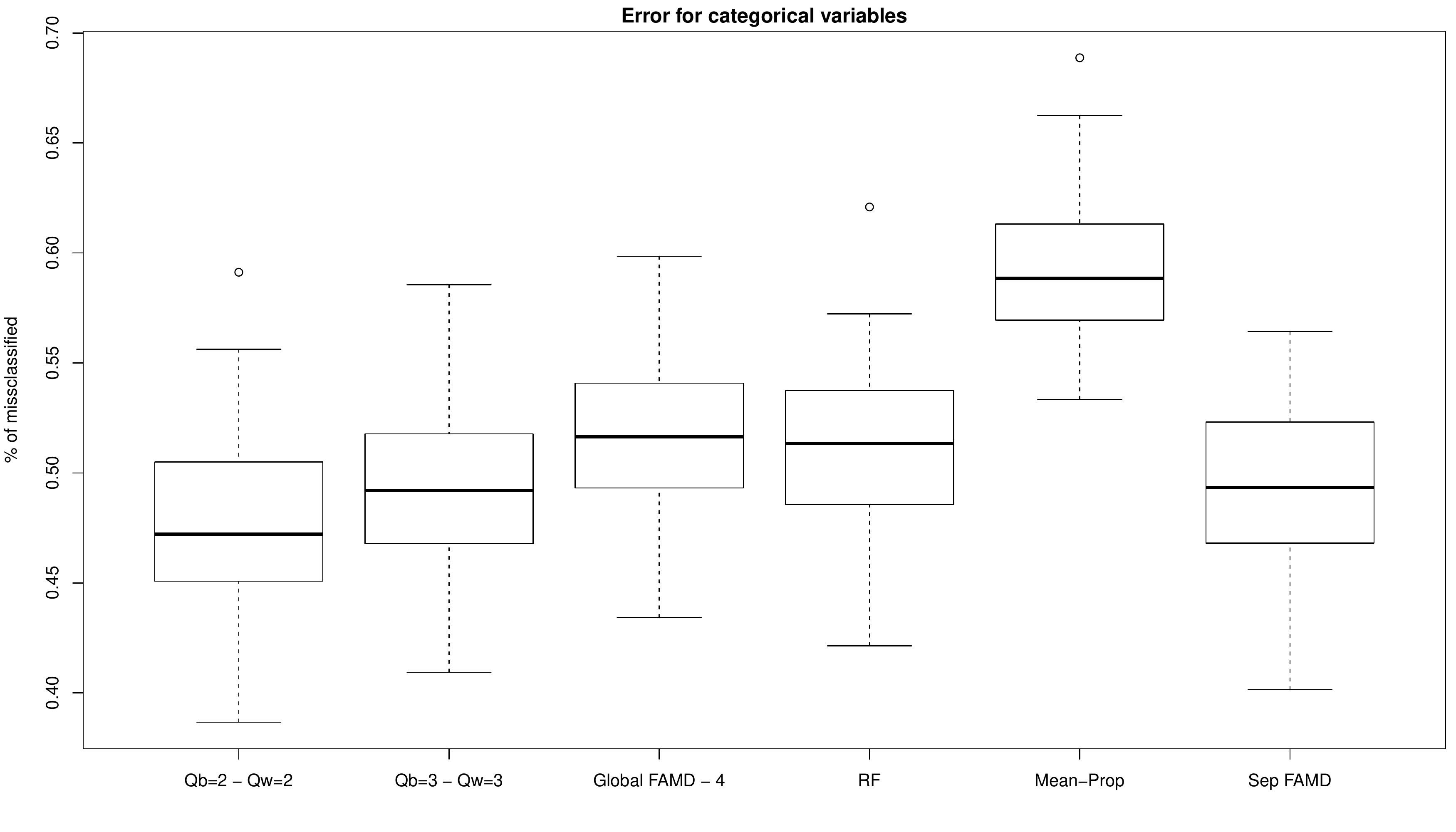} \caption{Data set with $J=10$ variables, 5 quantitative and 5 categorical variables, 20\% of missing values, $K=5$  groups and $n_k=30$ observations per group: on the left MSE for the quantitative variables; on the right percentage of misclassified for categorical variables. Multilevel FAMD is represented for different values of  $Q_b$ and $Q_w$ whereas Global FAMD with 4 dimensions. RF is the imputation with random forest, Mean-Prop means that the imputation is done by the mean for quantitative variables and the proportion for categorical ones, and sep FAMD gives the results when separate FAMD are performed on each group.
\label{fig:summaryR_right}}
\end{figure}
Figure \ref{fig:summaryR_right} shows again that imputing with the multilevel method gives better results than imputing with global FAMD or with random forests. This is especially true for the quantitative variables. Note that imputation with multilevel FAMD is quite stable with respect to the number of between and within components. 
As far as the computational time is concerned, we compare in Table \ref{tab:time} the performances of the different approaches. Regarding this point, SVD based imputation methods have a clear advantage over jomo and random forests.
\begin{table}
\center{
\begin{tabular}{|l | c c | c c|}
\hline
& $J=10$& $J=30$ &$J=15$& $J=35$\\ \hline
Global PCA &0.09& 0.3& & \\
jomo &11& 282 & & \\
Multilevel FAMD &1.5&1.2 & 2&  7 \\
Global FAMD&0.4&0.7&1& 4\\
Random forest &59&200&27&  246\\\hline
\end{tabular}
}
\caption{Time in seconds for a dataset with $20\%$ of missing values, $K=5$ groups and $n_k=200$ observations per groups, with 10 and 30 quantitative variables for the two left columns and with additional 5 categorical variables for the two right columns.\label{tab:time} }
\end{table}

\section{Hospital data analysis}
\label{trauma}
\subsection{Traumabase}
Our work is motivated by an application in public health on polytraumatized patients for 
the Traumabase\footnote{\url{http://www.traumabase.eu/fr_FR}} group
 at APHP (Public Assistance - Hospitals of Paris). 
Effective and timely management of major trauma patients is critical to improve outcomes and survival, given the high risks for the patient in case of delays or errors.
With the perspective of improving the decision-making process and the care of patients,
8 French Trauma centers have decided to collaborate to collect detailed high quality clinical
data from the scene of the accident to the exit of the hospital. The resulting database,
the Traumabase, has up to now gathered more than  7495 trauma
admissions data, and is permanently updated. \\
The data are highly heterogeneous, multi-source, and contain many missing values.
Furthermore, experts expect hospitals to have an influence on some of the variables, due to lack of practice standardization, and because the patients and their social status differ from one hospital to the other.  We analyse a portion of the initial data set containing $8$ features identified by physicians as prone to hospital effects. The data set of interest therefore consists in 5 qualitative and 3 quantitative variables measured over 7495 patients, and contains around 11\% of missing values; furthermore, there is at least one missing entry for 49\% of patients. There are certainly different generation mechanisms at work: some variables (such as the type of accident and the hospital center) are completely observed whereas the patterns of missingness of other variables (such as pelvic and lung X-ray) are believed to depend on the hospital center. In first approximation, a Missing At Random (MAR) mechanism - where the probability of missingness is allowed to depend on the observed variables - seems satisfying.\\

We focus on imputing of the Traumabase data with iterative MLFAMD with two aims. First, the imputed data can be further analyzed with other statistical methods such as predictive models, to predict some outcome of interest. However, care must be taken when analysing an imputed data set, as discussed in \Cref{conclusion}. 
Secondly, the imputation of missing data from a hospital is improved when the hospital is integrated into the aggregated database. Therefore, this may encourage them to share their data and participate in the medical data aggregation project. Such a project is important because
having at disposal aggregated data is an opportunity to have more patients and to develop more relevant modelling. Imputation is thus an incentive for hospitals to share their data and potentially lead to better care for all patients.\\
However, there are technical and social barriers to the aggregation of medical data. The size of combined databases often makes computations and storage intractable, while institutions are usually reluctant to share their data due to privacy concerns and proprietary attitudes. Both obstacles can be overcome by turning to distributed computations, which consists in leaving the data on sites and distributing the calculations, so that hospitals only share some intermediate  results instead of the raw data \citep{Naras17}. Among other methods, SVD, which only involves inner products and sums, can be very straightforwardly implemented in a distributed manner. Consequently, one main advantage of the methods we present is that they can also be distributed across sites. The distributed  framework is presented in  \Cref{sec:appDistSVD}.

\subsection{Simulated imputation of the Traumabase}

\begin{figure}[hbtp]  
\centering
 \includegraphics[width=7cm]{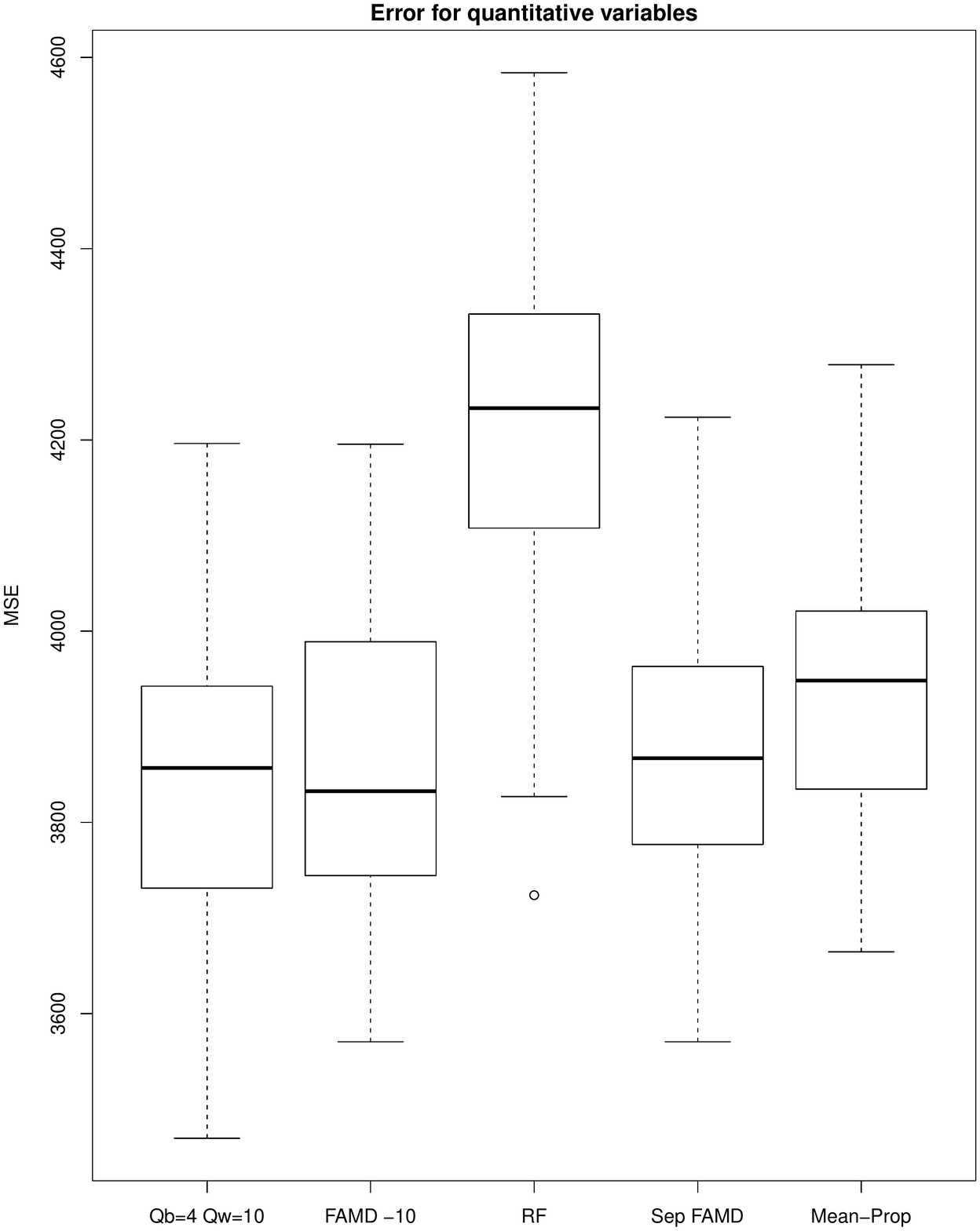} 
 \quad
 \includegraphics[width=7cm]{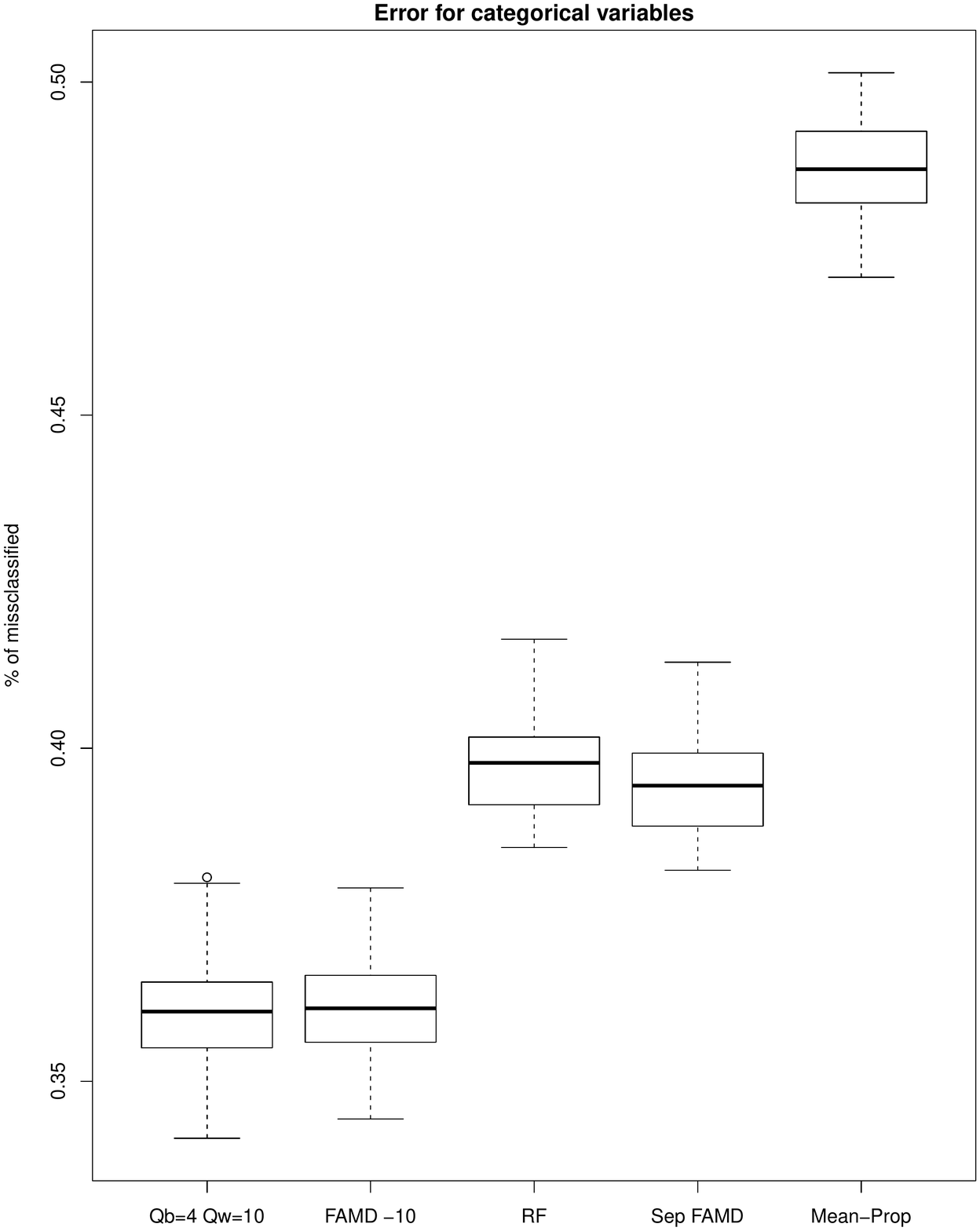}
 \caption{Traumabase: MSE of prediction and \% of mis-classification.\label{fig:hosp}}
\end{figure}

To assess the quality of imputation and legitimate the use of iterative MLFAMD to impute the Traumabase, we first perform simulations by inserting an additional of 10\% of missing values to the data set, predicting them with the different imputation methods described in \Cref{simu}, and computing the mean squared error of prediction for quantitative variables and the percentage of misclassification for categorical variables. Figure \ref{fig:hosp} presents the  results over 100 replications of the experiment.
In terms of prediction of quantitative variables, multilevel FAMD and global FAMD perform similarly and improve on the random forest imputation. We observe the same behavior for the categorical variables, with multilevel FAMD improving only slightly on global FAMD. Note that the data are quite difficult to impute and the relationship between variables weak.

\section{Conclusion}
\label{conclusion}

We proposed a method dedicated to the imputation of multilevel mixed data based on an iterative SVD algorithm. To the best of our knowledge this is the first multilevel method available for mixed data. 
Directions of future research include the development of an automated method to estimate the number of components $Q_b$ and $Q_w$. A first approach is for now to select $Q_b$ and $Q_w$ with cross-validation.
We are also eager to investigate a multiple imputation \citep{Murr}  procedure based on this multilevel component method, in order to further analyse the Traumabase data set with predictive models, for instance to study the occurence of diagnosis errors based on patients profiles. Multiple imputation is important to reflect the uncertainty associated to the imputed values.  We also believe the multilevel methods we have developped for mixed data can be useful for exploratory analysis and visualization.\\
Finally, as discussed, the methods presented in this paper can be implemented in parallel across groups or sites. A following project we are currently involved in consists in exploiting this property to implement a real-time distributed and privacy preserving platform, dedicated to the imputation of health care data partitioned across several hospitals, without having to aggregate the data. One issue with the distribution technique described in \Cref{sec:appDistSVD} is that we use iterative procedures, therefore after $N$ iterations each hospital has shared $N$ summary statistics, which can lead to information leakage. A possible solution to this problem is to resort to homomorphic encryption \citep{homenc} which allows to perform computations on encrypted data.

\section{Acknowledgement}
Genevi\'eve Robin was funded by the France-Stanford Center for Interdisciplinary Studies (visiting student researcher fellowship) to visit Stanford. The authors are grateful to the Traumabase group for providing the data and for their help.

\bibliographystyle{Chicago}
\bibliography{references}

\section{Appendix}

In this appendix, we show how to distribute multilevel iterative imputation algorithms in order to leave the data of each group on each site while applying the method. 

\subsection{Distributed rank-$Q$ PCA}
\label{sec:dist-pca}
We start by reminding the power method \citep{Golub:1996:MC:248979}, which computes the first left and right singular vectors of a matrix $Y\in \mathbb{R}^{n\times p}$. Without loss of generality, we assume $n\leq p$. Suppose $Y = U\Lambda^{1/2} V^{\top}$, $U = \left(u_1,\ldots, u_n\right)$, $V = \left(v_1,\ldots, v_n\right)$ and $\Lambda = \operatorname{diag}(\lambda_1^2,\ldots,\lambda_n^2)$ $|\lambda_1|\geq|\lambda_2|\ldots\geq|\lambda_n|$. The power method is iterative and produces sequences of vectors $z^{(t)}$ and $q^{(t)}$ converging to $u_1$ and $v_1$ respectively, with iterations detailed in Algorithm \ref{power-method}. Let $q^{(0)}$ be a starting point satisfying $\norm{q^{(0)}}[2] =1$.
\begin{figure}[hbtp]
  \centering
  \begin{minipage}{.7\linewidth}
    \begin{algorithm}[H]
      \SetAlgoLined
      \For{$t=1,2,\ldots$}{
        $z^{(t)} = Y^{\top}q^{(t-1)}$\\
        $z^{(t)} = z^{(t)}/\Norm{z^{(t)}}_2$\\
        $q^{(t)} = Yz^{(t)}$\\
        $\lambda^{(k)} = \Norm{q^{(t)}}_2$\\
        $q^{(t)} = q^{(t)}/\Norm{q^{(t)}}_2$
      }
      \caption{Power method}
      \label{power-method}
    \end{algorithm}
  \end{minipage}
\end{figure}
The sequences $q^{(t)}$ and $z^{(t)}$ converge to $u_1$ and $v_1$ respectively, when $\pscal{q^{(0)}}{u_1}\neq 0$ and $|\lambda_1|>|\lambda_2|$; the rate of convergence is dictated by the ratio $|\lambda_2|/|\lambda_1|$. This directly extends to the computation of the rank-$Q$ SVD. One can actually estimate $u_1$, $v_1$ and $\lambda_1$, then the second dimension by applying the same procedure to $Y-u_1\lambda_1v_1^{\top}$, and so on so forth. Moreover it is straightforward to distribute this procedure when the data are grouped in $K$ different sites with 
$$Y = \left(
\begin{array}{c}
Y_1 \\
\hline
Y_2\\
\hline
\vdots\\
\hline
Y_K
\end{array}
\right) .$$
Indeed, all the computations in Algorithm \ref{power-method} can be done in parallel with a master-slave architecture \citep{Naras17},  where a central server collects summary statistics computed locally on sites, as illustrated Figure \ref{marg-tree}.
Here,  the local right singular vectors $v_j$, $j\in\{1,\ldots,n\}$ are sent to the master.
The corresponding algorithm is given in Algorithm \ref{dist-power-method}, and leads exactly to applying the power method for rank-$Q$ SVD to the entire data matrix $Y$. The procedure is implemented in the distcomp R package \citep{Naras17}.
\begin{figure}[hbtp]
  \centering
  \footnotesize
  \begin{minipage}{.7\linewidth}
    \begin{algorithm}[H]
      \SetAlgoLined
      \KwData{Workers private data $Y_k\in \mathbb{R}^{n_k\times p}$}
      \KwResult{$F\in\mathbb{R}^{n\times Q}$, $V\in\mathbb{R}^{p\times Q}$, $\lambda_1\geq \lambda_2\geq \ldots\geq \lambda_Q$}
      $F = 0$, $\lambda = 0$\\
      \For{$k = 1,\ldots K$}{
        $F_k = 0$\\
        transmit $n_k$ to master\
      }
      \For{$i=1,\ldots,Q$}{
        \For{$k = 1,\ldots K$}{
          $q_k = (1,1,\ldots, 1)$\
      }
      	$\Norm{q}_2 = \sqrt{\sum_{k=1}^K n_k}$\\
      	transmit $\Norm{q}_2$, $V$ and $\lambda$ to workers\\
      	\Repeat{convergence}{
      		\For{$k=1,\ldots,K$}{
				$q_{k} = q^{k}/\Norm{q}_2$\\
				$r_{k} = (Y_k - F_kV^{\top})^{\top}q_{k}$\\
				transmit $r_k$ to master\
			}
			$r = \sum_{k=1}^K r_k$\\
			$r = r/\Norm{r}_2$\\
			transmit $r$ to workers\
			\For{$k=1,\ldots, K$}{
				$q_k = Y_kr$\\
				transmit $\Norm{q_k}_2$ to master\
			}
			$\Norm{q}_2 = \sum_{k=1}^K\Norm{q_k}_2$\\
			transmit $\Norm{q}_2$ to workers\\
			$\lambda_i = \Norm{q}_2$\
      	}
      	$V = cbind(V, r)$\\
		\For{$k=1,\ldots,K$}{
			$F_k = \text{combine by column }(F_k, q_k)$\
		}
      }
      \caption{Distributed power method}
      \label{dist-power-method}
    \end{algorithm}
  \end{minipage}
\end{figure}

\subsection{Distributed algorithm for iterative multilevel PCA}
\label{sec:appDistSVD}

\tikzstyle{level 1}=[level distance=5cm, sibling distance=4cm]
\tikzstyle{level 2}=[level distance=5cm, sibling distance=4cm]

\tikzstyle{bag} = [text width=4em, text centered]
\tikzstyle{end} = [circle, minimum width=3pt,fill, inner sep=0pt]

\begin{figure}[hbtp]
\centering
\begin{tikzpicture}[grow=right, sloped]
\node[bag] {Master}
    child {
        node[bag] {Hospital 3}        
            edge from parent 
            node[above] {sends $m_3, \pi_3, V_{w,3}, n_3$}
            node[below]  {receives $m, \pi, V_w$}
    }
    child {
        node[bag] {Hospital 2}        
        edge from parent         
            node[above] {sends $m_2, \pi_2, V_{w,2}, n_2$}
            node[below]  {receives $m, \pi, V_w$}
    }
    child {
        node[bag] {Hospital 1}        
        edge from parent         
            node[above] {sends $m_1, \pi_1, V_{w,1}, n_1$}
            node[below]  {receives $m, \pi, V_w$}
    };
\end{tikzpicture}
\caption{Master-slave distribution structure. The hospitals send their local means, proportions, sample size and right sigular vectors to the master. The master sends back the overall means, proportions, and right singular vectors to the hospitals.}
\label{marg-tree}
\end{figure}
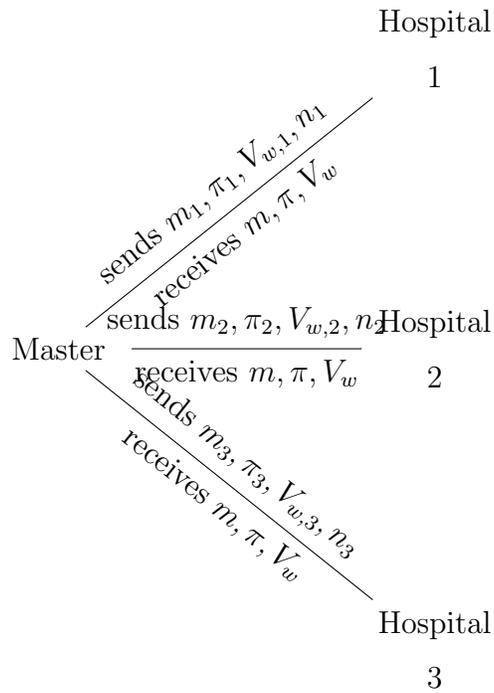
In \Cref{sec:dist-pca}, we see how the power method \citep{Golub:1996:MC:248979}, which computes the first left and right singular vectors of a matrix $Y\in \mathbb{R}^{n\times p}$, can be straightforwardly distributed over $K$ different sites. This algorithm can then be used to perform a distributed rank-$Q$ SVD, as shown in Algorithm \ref{dist-power-method}. We take advantage of this property to develop a distributed version of the iterative PCA algorithm, presented in Algorithm \ref{algo:multi-pca}. This algorithm imputes missing values with the iterative PCA algorithm in a distributed way. Indeed, iterative PCA imputation involves iterative SVD. Plugged in Algorithm \ref{algo:it-PCA}, Algorithm \ref{algo:multi-pca} leads to a distributed version of the iterative multilevel PCA algorithm. In the same way, distributed iterative MLMCA and MLFAMD are implemented.

\begin{figure}[hbtp]
  \centering
  \footnotesize
  \begin{minipage}{.7\linewidth}
\begin{algorithm}[H]
\SetAlgoLined
\KwData{$Y_k\in \mathbb{R}^{n_k\times p}$, $Q_b$, $Q_w$}
\KwResult{$\hat{m}$, $F_{b}$,$V_b$,$F_{w}$,$V_{w}$}
\textbf{Initialization}: impute missing values with initial values; $(n\times p)=diag(\sqrt{n_k})$.\;
$R = 0$, $\lambda = 0$\\
\For{$k = 1,\ldots K$}{
	$F_k = 0$\\
	transmit $n_k$ to master\
}
\For{$i=1,\ldots,Q$}{
	\For{$k=1,\ldots,K$}{
		$q_k = (1,1,\ldots, 1)$\
	}
	$\Norm{q}_2 = \sqrt{\sum_{k=1}^K n_k}$\\
	transmit $\Norm{q}_2$, $V$ and $\lambda$ to workers\\
	\Repeat{convergence}{
		\For{$k=1,\ldots,K$}{
			$q_k = q_k/\Norm{q}_2$\\
			$r_k = (Y_k - F_kV^{\top})^{\top}q_k$\\
			transmit $r_k$ to master
		}
		$r = \sum_{k=1}^K r_k$
		$r = r/\Norm{r}_2$
		transmit $r$ to workers\\
		\For{$k=1,\ldots, K$}{
			$q_k = Y_kr$\\
			transmit $\Norm{q_k}_2$ to master
		}
		$\Norm{q}_2 = \sum_{k=1}^K\Norm{q_k}_2$\\
		transmit $\Norm{q}_2$ to workers\\
		$\lambda_i = \Norm{q}_2$
	}
	$V = \text{combine by column }(V, \sqrt{\lambda_i}r)$\\
	\For{$k=1,\ldots,K$}{
		$F_k = \text{combine by column }(F_k, q_k)$
	}
}
\caption{Distributed iterative PCA}
\label{algo:multi-pca}
\end{algorithm}
\end{minipage}
\end{figure}

\end{document}